\documentclass[aps,prl,reprint,floatfix,showpacs]{revtex4-1}

 \usepackage{amssymb}
 \usepackage{amsmath}
 \usepackage{amsfonts}
 \usepackage{graphicx}

\usepackage{pdfpages}
\makeatletter
\AtBeginDocument{\let\LS@rot\@undefined}
\makeatother

 \newcommand{\fancy}{\mathcal}

 \newcommand{\FE}{\kappa}
 \newcommand{\FC}{\fancy{C}}

 \newcommand{\FG}{\fancy{G}}

 \newcommand{\lbrs}{\left[}
 \newcommand{\rbrs}{\right]}

 \renewcommand{\vec}[1]{\boldsymbol{#1}}
 
 \newcommand{\mat}[1]{\mathbb{#1}}

 \newcommand{\beq}{\begin{eqnarray}}
 \newcommand{\eeq}{\end{eqnarray}}
 
 \newcommand{\tr}{\text{Tr}}
 \newcommand{\Tr}[1]{{{\text{Tr}}\lbrs #1 \rbrs}}

 \newcommand{\half}{\frac{1}{2}}

 \newcommand{\rcite}[1]{Ref.~\onlinecite{#1}}

 \newcommand{\rcites}[1]{Refs~\onlinecite{#1}}

 \newcommand{\Hx}{\text{Hx}}

 \newcommand{\crm}{\text{c}}

 \newcommand{\Hxc}{\text{Hxc}}

 \newcommand{\EHxc}{{\cal E}_{\Hxc}}
 \newcommand{\EHx}{{\cal E}_{\Hx}}
 \newcommand{\Ec}{{\cal E}_{\crm}}
 
 \newcommand{\Ts}{{\cal T}_{s}}
 
 \newcommand{\E}{{\cal E}}
 \newcommand{\F}{{\cal F}}

 \newcommand{\mU}{\mat{U}}

 \newcommand{\pr}{^{\prime}}

 \renewcommand{\vr}{\vec{r}}
 
 \newcommand{\vrp}{\vec{r}\pr}

 \newcommand{\ibra}[1]{\langle#1|}
 \newcommand{\iket}[1]{|#1\rangle}
 
 \newcommand{\ibraketop}[3]{\langle#1|#2|#3\rangle}
 \newcommand{\ibkouter}[1]{|#1\rangle\langle#1|}
 
 \newcommand{\iout}{\ibkouter}

 \newcommand{\ieri}[2]{(#1|#2)}

 \newcommand{\Ext}{{\text{Ext}}}

 \newcommand{\up}{\mathord{\uparrow}}
 \newcommand{\down}{\mathord{\downarrow}}

 \newcommand{\nh}{\hat{n}}
 \newcommand{\Th}{\hat{T}}
 
 \newcommand{\Wh}{\hat{W}}
 \newcommand{\vh}{\hat{v}}
 
 \newcommand{\Hh}{\hat{H}}

 \newcommand{\rhoh}{\hat{\rho}}
 
 \newcommand{\Gammah}{\hat{\Gamma}}

 \newcommand{\Gammahmol}{\hat{\Gamma}_{\text{mol}}}
 \newcommand{\Gammahloc}{\hat{\Gamma}_{\text{loc}}}

 \usepackage[normalem]{ulem}
 \usepackage{color}
 \usepackage{xcolor}

 \definecolor{Mygrey}{gray}{0.80}

 \newcommand*{\LightComments}{}%

 \ifdefined\NoComments

 \else
  \ifdefined\LightComments

  \else

  \fi
 \fi

\begin{document}
 \title{`Hartree-exchange' in ensemble density functional theory: \\
 Avoiding the non-uniqueness disaster}
 \author{Tim Gould}\affiliation{Qld Micro- and Nanotechnology Centre, %
   Griffith University, Nathan, Qld 4111, Australia}
 \author{Stefano Pittalis}\affiliation{CNR-Istituto di Nanoscienze, Via
   Campi 213A, I-41125 Modena, Italy}
 \begin{abstract}
   Ensemble density functional theory is a promising method for the
   efficient and accurate calculation of excitations of quantum
   systems, at least if useful functionals can be developed to
   broaden its domain of practical applicability.
   Here, we introduce a guaranteed single-valued
   `Hartree-exchange' ensemble density functional,
   $\EHx[n]$, in terms of the right derivative of the universal
   ensemble density functional with respect to the coupling constant at
   vanishing interaction.
   We show that $\EHx[n]$ is straightforwardly
   expressible using block eigenvalues of a simple matrix
   [equation~(14)].
   Specialized expressions for $\EHx[n]$ from the literature,
   including those involving superpositions of Slater
   determinants, can now be regarded as originating from the unifying
   picture presented here.
   We thus establish a clear and practical description for
   Hartree-exchange in ensemble systems.
 \end{abstract}
 \maketitle

 Density functional theory\cite{HohenbergKohn,KohnSham} (DFT) is,
 arguably, the most important methodology in electronic structure
 theory due to its remarkable accuracy in numerically efficient
 approximations. But ``open'' systems that mix different numbers of
 electrons, degenerate groundstates, and excited states have long posed
 a challenge to conventional approaches [see
   e.g. \rcites{Yang2000,MoriSanchez2006,Cohen2008,MoriSanchez2009,%
     Savin2009-Size,Cohen2012-Challenges,Levy2014}], and can make even
 qualitative accuracy very difficult to achieve. One promising route
 around these problems is to employ ensemble density functional theory
 \cite{Valone1980,Perdew1982,Lieb1983,Savin1996,Ayers2006-Axiomatic,%
   Gross1988-1,Gross1988-2,Oliveira1988-3} (EDFT), in which ensembles
 of quantum states extend the original pure state approach of DFT to
 such systems. As many quantum systems\cite{Gaitan2009,Malet2015}
 are better understood by models involving ensembles, ideas
 and constructions at the heart of EDFT offer a more
 promising approach for their study, compared to conventional DFT.
 The ability to use EDFT as successfully and
 easily as we now use DFT could thus
 transform quantitative understanding of numerous quantum systems
 and processes, such as charge transfer and diabatic reactions.

 In standard DFT, we decompose
 the universal functional, $F$, of the particle density $n$, as
 \begin{align}\label{FDFT}
   F[n] &= T_s[n] + E_{\Hx}[n] + E_{\crm}[n]
 \end{align}
 where $T_s$ is the kinetic-energy density of the Kohn-Sham (KS)
 reference system,
 \begin{align}
   E_{\Hx}[n]  &= \int\frac{d\vr d\vrp}{2|\vr-\vrp|}
   \bigg\{
   n(\vr)n(\vr') - \big| \rho_s(\vr,\vrp) \big|^2 \bigg\}
   \;,
   \label{eqn:EHxSS}
 \end{align}
 is the Hartree energy plus the exchange energy -- in which
 $\rho_s(\vr,\vr')$ is the KS one-body reduced density matrix and
 $n(\vr) = \rho_s(\vr,\vr)$ equals the interacting ground state
 particle density -- and $E_c[n]$ is the correlation energy.

 It may be tempting to switch to EDFT by replacing the pure state
 quantities with ensembles (statistical mixtures of pure states)
 by performing a simple replacement of the
 particle density by its ensemble generalization.
 We thus set $\rho_s(\vr,\vr') \to
 \tr[\Gammah^{n}_0\hat{\rho}(\vr,\vr')]$ where
 $\Gammah^{n}_0$ is the ``ensemble density matrix'' operator
 describing the reference Kohn-Sham state,
 and use $\tr[\Gammah^{n}_0\hat{\rho}(\vr,\vr)]=n(\vr)$
 to write
 \begin{align*}
   E_{\Hx}[n]   & \rightarrow \int \frac{d\vr d\vrp}{2|\vr-\vrp|}
   \bigg\{ n(\vr)n(\vrp)
     -\big|\tr[\Gammah_0^n\rhoh(\vr,\vr')]\big|^2 \bigg\}
     \;.
 \end{align*}
 This, however, comes at the price of introducing spurious
 ``ghost interactions'' to both the Hartree and exchange terms --
 with sometimes disastrous consequences in approximate
 calculation\cite{Gidopoulos2002}. 

 A ghost interaction error can be understood as a generalization of the
 one- or $N$-particle self-interaction error\cite{MoriSanchez2006}.
 But, rather than an orbital spuriously interacting with itself it
 instead spuriously interacts with its `ghost' counterpart in a
 \emph{different} replica of the same system.
 From this understanding comes a desire to correct these ghost
 interactions in the Hartree and exchange energies.
 A formal justification of these corrections was put forward by
 Gidopoulos \emph{et al}\cite{Gidopoulos2002} by importing the result for
 the Hartree-Fock approximations for ensembles -- which was noticed to
 be ghost-interaction free for the cases considered in the same
 cited work -- and then invoking an extended optimized effective
 potential method\cite{OEP1,OEP2,Nagy1998}, as in pure state exchange theory.

 The principle espoused by Gidopoulos \emph{et al} is clear.
 Unpleasantly, however,
 the resulting prescription must be worked out for each case at
 hand. This process entails rather tedious and system specific
 bookkeeping. For specific cases, the expression of the
 resulting functional may be found in the original literature%
 \cite{Gidopoulos2002,Cohen2008,Gould2013-LEXX,Pastorczak2013-Multi}. 

 Matters are simplified somewhat by working with
 combined `Hartree-exchange' (Hx) expressions  to avoid
 difficulties of treating the components individually as in 
 $\tr[\Gammah_s \Wh]$ where $\Wh$ is the electron-electron interacting
 operator. This approach has been elaborated further in Nagy's
 works\cite{Nagy1995-exact,Nagy1998}
 showing how ghost-interaction free Hartree plus exchange-correlation
 functionals can be defined in terms of  weighted sum of energies of
 the states in the ensembles.
 Such an approach resolves many problems and is, indeed, correct in
 many instances. Very recently, the evolution of `Hartree-exchange' in
 EDFT has culminated with the proposal of a Symmetrized Eigenstate
 Hartree-exchange expression
 (SEHX)~\cite{Yang2017-EDFT,Pribram-Jones2014,Yang2014}. 

 In this work, we put forward a universal and unifying treatment of the
 `Hartree-exchange' in EDFT that encompasses other specialized
 `Hartree-exchange' expressions -- when appropriate.
 We start from the observation that using $\tr[\Gammah \Wh]$ too
 directly can lead to subtle fundamental issues.
 As we will illustrate, this occurs in the presence of particular forms
 of degeneracies in the many-body quantum states for which
 $\tr[\Gammah\Wh]$ is not characterized by a unique value,
   as happens when the same ensemble particle density may be obtained
   from different $\Gammah$. In such cases the 
 possibility of extracting a simple \emph{single valued} expression for
 a corresponding Hartree-exchange ensemble density functional
 is far from obvious.
 The goal of this work is to avoid such a
 ``non-uniqueness disaster'', and thus develop a functional which
 can be used directly or approximated for calculations.

 We will show that a general approach free from pathologies
 follows directly from a definition of `Hartree-exchange' that
 avoids a direct reference to the Hartree-Fock method
 by working consistently within EDFT. The core of the idea is to
 exploit the ``nice'' properties
 of the \emph{exact} universal interacting functionals, $\F^{\lambda}[n]$
 [see eq.~\eqref{eqn:Flambda} and prior discussion],
 and its Kohn-Sham version, the kinetic energy density
 functional $\Ts[n]\equiv \F^0[n]$, by defining
 \begin{align}
   \EHx[n] := \lim_{\lambda \to 0^+}\frac{\F^{\lambda}[n]-\Ts[n]}{\lambda}
   \label{eqn:EHxLim}
 \end{align}
 where $\lambda$ stands for the \emph{continuously} varying strength of
 the electron-electron interaction coupling operator $\Wh$. 
 Obviously, the companion correlation functional for
 $\EHx[n]$ must be $\Ec[n] := \F^1[n] - \Ts[n] - \EHx[n]$.

 As we shall illustrate, Eq.~\eqref{eqn:EHxLim} has several convenient
 corollaries
 [Eqs~\eqref{eqn:EHxLim3},~\eqref{eqn:EHxLimU},~\eqref{eqn:EHxLimEig}].
 For now, we highlight that it can be restated in terms of a minimization
 \begin{align}\label{eqn:EHxMin}
   \EHx[n] \equiv \min_{ \Gammah \in \FG^{n,\lambda = 0}}
   \Tr{\Gammah\Wh}\;,
 \end{align}
 where $ \FG^{n,\lambda = 0}$ is the set of the non-interacting 
 ensembles which yield the prescribed particle density $n$ and the
 exact Kohn-Sham kinetic energy $\Ts[n]$. Crucially, here the
 conventional restriction to single Slater determinant must be
 avoided, and we must concern ourselves only with the
 uniqueness of $\EHx$ and not the states $\Gammah$ that yield it.

 From \eqref{eqn:EHxLim} and \eqref{eqn:EHxMin}
 we can gain several important insights of both formal and
 practical use (in the remainder of the paper, we will come back on the
 following points by providing further details or explicit examples):

 \emph{Avoidance of the non-uniqueness disaster}:
 $\tr[\Gammah \Wh]$ can assume different values on different states
 $\Gammah$ having equal kinetic energy and equal densities.
 In this situation, a functional expression directly based on
 $\tr[\Gammah \Wh]$ does  not provide  a unique value for a given
 density! $\EHx[n]$, instead, picks automatically the minimum value. In
 other words, $\EHx[n]$ is guaranteed to acquire unique values for
 each given $n$, regardless of any non-uniqeness of
 $\Gammah \rightarrow n$;

 \emph{Maximal freedom from interactions}:
 It is reassuring to note that avoiding the non-uniquness disaster does
 not play against avoiding the spurious interactions we discussed
 above. This readily follow from the fact that $\EHx[n]$ is determined
 by a \emph{minimization} of the expectation value of the interaction
 energy. Therefore, it must also be maximally free from \emph{spurious}
 interactions -- as much as the constraints underlying the construction
 of $\FG^{n,\lambda = 0}$ can allow;

 \emph{Multireference states}:
 It also becomes apparent that all the ``nice'' properties of $\EHx[n]$
 induced by $\F^{\lambda}[n]$ can be spoiled by restrictions on the
 underlying admissible ensembles.  In some situations, described
 later, insistence on using single Slater determinants, as in
 conventional Kohn-Sham theory, is overly restrictive in EDFT.

 In the remainder of this paper we shall first explore the derivation
 of, and consequences of this unifying picture of Hartree-exchange
 physics. Finally we will conclude.


 Let us first consider some background theory, and define some key
 concepts.
 Ensemble DFT \cite{Valone1980,Perdew1982,Lieb1983,Ayers2006-Axiomatic}
 overcomes the restriction of standard Kohn-Sham theory to pure
 states. Thus, it can be used to widely generalize the systems that can
 be studied using DFT.
 {\em Firstly}, we can form ensembles
 $\Gammah=\sum_{\FE}w_{\FE}\iout{\FE}$ that mix degenerate states, such
 that $\tr[\Hh\Gammah^{\text{degen}}]=\sum_{\FE}w_{\FE}E_0=E_0$
 for the appropriate Hamiltonian $\Hh$.
 {\em Secondly}, we can form ensembles that involve excited states,
 provided the weights do not increase as the energy increases.
 Thus,
 $\tr[\Hh\Gammah^{\text{excite}}]=\sum_{\FE}w_{\FE}E_{\FE}\geq E_0$,
 and the energy can be higher than the equivalent groundstate
 $E_0=\ibraketop{0}{\Hh}{0}$. The degenerate case above is,
 in fact, an important specialized version of this type of ensemble.
 For the sake of simplicity, we shall not consider ensembles that mix
 different numbers of electrons (``open'' states).

 In order to account for the aforementioned cases, some
 constraints -- which we indicate collectively as $\FC$ -- must be
 enforced to suit the problem e.g.
 to select certain fractions of ground- and excited states or
 to preserve certain symmetries on the system.
 Moreover, since Eq.~\eqref{eqn:EHxLim} invokes the use of the
 adiabatic coupling, they must define weights at varying electron
 interaction strength $\lambda$ that reproduce the particle 
 density at full interaction, a non-trivial task.
 The present work relies on two primary, physically-reasonable
 restrictions on the allowed ensembles:
 i) they must describe a \emph{finite} number of `typical' states with
 weights that can change only at an energy-level crossing so that
 $\exists \delta>0$ where
 $w_{\FE}^{\lambda}=w_{\FE}^0,~\forall~0\leq\lambda<\delta$;
 ii) the density $n$ must be ensemble $v$-representible
 for $0\leq\lambda<\delta$ so that we can invoke
 the theorems of Gross, Oliveira and Kohn%
 \cite{Gross1988-1,Gross1988-2,Oliveira1988-3}
 to define $\F^{\lambda}[n]$.

 In the parlance of the constrained search approach, one can then
 define a general universal functional
 \begin{align}
  \F^{\lambda}_{\FC}[n]=&
  \min_{\Gammah_{\FC} \to n}\Tr{ \Gammah(\Th + \lambda\Wh) }
  \label{eqn:Flambda}
 \end{align}
 where the minimization (strictly an infimum) is carried out over
 density matrices $\Gammah_{\FC}$ constrained to give a density
 $n=\tr[\Gammah_{\FC}\nh]$ by means of orthonormal states and,
 a prescribed set of weights $w_{\FE}^{\lambda}$.
 $\Th$ is a kinetic energy operator and 
 $\Wh$ is a positive-definite two-point interaction operator.
 For what follows, however, we only need
 to be concerned with $0\leq \lambda<\delta$ and $\lambda = 1$.

 The universal functional \eqref{eqn:Flambda} then allows us to find the
 energy and density of any Hamiltonian
 $\{ \Th + \Wh + \vh_{\Ext} \}$ by seeking
 \begin{align}
  \E[v_{\Ext}]=&\min_{n\to N}\bigg\{ \F[n] + \int d\vr
  n(\vr)v_{\Ext}(\vr) \bigg\},
 \end{align}
 where $\F[n] = \F^1[n] = \Ts[n] + \EHxc[n] $ and
 $v_{\Ext}(\vr)$ is an external (local) multiplicative potential.
 We use $\F^{\lambda}$ to define the Kohn-Sham kinetic energy
 $\Ts[n]=\F^0[n]$ and Hartree, exchange and correlation (Hxc)
 energy $\EHxc[n]=\F[n]- \Ts[n]$ at given density $n$.
 Here and henceforth $\FC$ is left off for succinctness,
 unless strictly necessary to improve clarity.

 Now, let us analyze the Hx energy definition
 \begin{align}
  \EHx[n] := \lim_{\lambda \to 0^+}
  \frac{\F^{\lambda}[n]-\F^0[n]}{\lambda}
  \equiv \partial_\lambda \F^\lambda[n]{\Big |}_{\lambda= 0^+} \;.
  \label{eqn:EHxLim2}
 \end{align}
 First, let us assume that there is a unique interacting ensemble
 $\Gammah^{n,\lambda}$, which is connected perturbatively to a
 non-interacting one $\Gammah^{n,0}$ along the adiabatic
 connection. Consider the case for which non-degenerate perturbation
 theory applies (the degenerate case is considered below).
 For $\lambda \ll 1$, we can write
 \begin{align} 
  \Gammah^{n,\lambda}\approx \Gammah^{n,0}+\lambda\Delta\Gammah
  =\sum_{\FE}w_{\FE}\iout{\FE;0} + O(\lambda)\;,
 \end{align}
 where $w_{\FE}\equiv w_{\FE}^0$ for $0\leq \lambda<\delta$. In this case,
 \begin{align}
  \F^{\lambda}[n]
  =&\Ts[n] + \lambda \tr[\Gammah^{n,0}\Wh] + O(\lambda^2).
  \label{eqn:FetaWb}
 \end{align}
 Here, we have used $\tr[\Gammah^{n,\lambda}\Th]=\Ts[n]+O(\lambda^2)$,
 which follows from rewriting $\Ts \leq \tr[\Gammah^{n,\lambda}\Th]$
 and $\F^{\lambda} \leq \tr[\Gammah^{n,0} (\Th+\lambda\Wh)]$
 [from \eqref{eqn:Flambda}] as
 $0\leq \tr[\Gammah^{n,\lambda}\Th]-\Ts\leq 
 \lambda\tr[(\Gammah^{n,0}-\Gammah^{n,\lambda})\Wh]$
 and letting $\lambda\to 0$.
 Thus,
 \begin{align}
  \EHx[n]= \lim_{\lambda \to 0^+}
  \frac{\F^{\lambda}[n]-\Ts[n]}{\lambda}
  = \tr[\Gammah^{n,0}\Wh]
  \label{eqn:EHxLim3}
 \end{align}
 follows straightforwardly.
 Clearly, when constraints $\FC$ allow only a pure state such that
 $\Gammah^{n,0}=\iout{\Phi_s}$, we readily get
 $$\EHx[n]=\ibraketop{\Phi_s}{\Wh}{\Phi_s}
 \equiv E_{\rm Hx}[n]$$ giving \eqref{eqn:EHxSS} for regular
 non-ensemble theory.
 If, instead, $\Gammah^{n,0} = w\sum_{\FE} \iout{\Phi_s^{\FE}}$
 for unique states $\iket{\Phi_s^{\FE}}$, as in the case of an
 equiensemble $w_{\FE} = w$, we end up with
 \begin{align}
  \EHx[n] = w \sum_{\FE} \ibraketop{\Phi_s^{\FE}}{\Wh}{\Phi_s^{\FE}}\;,
  \label{eqn:EHxMin2}
 \end{align} 
 where we purposely refrained from using common rewritings such as
 $\ibraketop{\Phi_s^{\FE}}{\Wh}{\Phi_s^{\FE}} = E_{\rm Hx}[ \Phi_s^{\FE} ] =
 E_{\rm Hx}[ n_s^{\FE} ] $ to avoid the risk of confusion.
 Ref.~\cite{Gould2013-LEXX} reports details of several specialized
 examples of this type, including equiensembles over \emph{all}
 members of a symmetry group.

 Next, let us consider the case where we must account for
 degeneracies. Thus, for $\lambda \ll 1$, we write
 \begin{align}
  \iket{\FE; \lambda}
  =&\sum_{\FE'}U_{\FE'\FE}\iket{\Phi^{\FE}} + O(\lambda)\;,
  \label{eqn:SDU}
 \end{align}
 in which superpositions of different non-interacting Slater
 determinants $\iket{\Phi^{\FE}}$ having the same kinetic energies
 and densities are allowed.
 The matrix $\mU$ is a unitary transformation that,
 regardless of the weights,
 leaves $\tr[\Gammah_{\mU}\Th]$ and $\tr[\Gammah_{\mU}\nh]$
 unchanged for $\Gammah_{\mU}\equiv \sum_{\FE} w_{\FE} \iout{\FE; \lambda}
 =\sum_{\FE\FE_1\FE_2} w_{\FE} U_{\FE_1\FE}U^*_{\FE_2\FE}
 \iket{\Phi^{\FE_1}}\ibra{\Phi^{\FE_2}}$, using \eqref{eqn:SDU}. However,
 $ \tr[\Gammah_{\mU}\Wh]=\sum_{\FE\FE_1\FE_2}
  w_{\FE}U_{\FE_1\FE}U^*_{\FE_2\FE}
  \ibraketop{\Phi^{ {\FE_1}}}{\Wh}{\Phi^{\FE_2}} + O(\lambda)$
 may change with $\mU$. From the minimization of
 $\tr[\Gammah^{n,\lambda}(\Th+\lambda\Wh)]$, we therefore get
 \begin{align}
  \EHx[n]=&
  \min_{\mU}\Tr{ \Gammah_{\mU} \Wh}\;,
  \label{eqn:EHxLimU}
 \end{align}
 which naturally incorporates multi-reference states, when appropriate
 (see the He example given just below).

 Finally, to accommodate both the non-degenerate and degenerate
 cases that can arise simultaneously in Gross, Oliveira and Kohn%
 \cite{Gross1988-1,Gross1988-2,Oliveira1988-3}
 ensembles, we write the most general and amenable formula
 \begin{align}
  \EHx[n]=&\sum_{\FE}w_{\FE}\Lambda_{\Hx,\FE}[n]\;.
  \label{eqn:EHxLimEig}
 \end{align}
 Here, $\Lambda_{\Hx,\FE}[n]$ are ``block eigenvalues''
 of $\mat{W}\equiv \ibraketop{\Phi^{ {\FE}}}{\Wh}{\Phi^{\FE'}}$, obtained
 by diagonalizing submatrices $\mat{W}^b$ (the blocks) of $\mat{W}$
 composed of states with the same kinetic energies and densities,
 and ordering the eigenvalues within each block whilst preserving
 the order of the blocks.
 Thus, non-degenerate states correspond to blocks of one element,
 and the Hx energy of degenerate states is guaranteed minimized by
 the ordered eigenvalues because of the Schur-Horn theorem.
 A detailed proof of \eqref{eqn:EHxLimEig} is provided in the
 Supplemental Material\cite{Supp}.

 By way of example, consider the He atom in an excited state ensemble
 composed of the groundstate $\iket{0}=\iket{1s\down 1s\up}$,
 and an arbitrary mix of the a fourfold
 degenerate set of lowest excited single Slater determinants
 $\iket{1}=\iket{1s\up 2s\up}$,
 $\iket{2}=\iket{1s\down 2s\down}$,
 $\iket{3}=\iket{1s\up 2s\down}$, 
 $\iket{4}=\iket{1s\down 2s\up}$.
 Previously reported work in
 \rcites{Pribram-Jones2014,Yang2014,Yang2017-EDFT} has identified and
 detailed how the correct spectra of the atom must be obtained
 by considering superpositions of Kohn-Sham determinants. 
 Our goal in the following is merely to show that our approach includes
 such a result directly.

 State $\iket{0}$ has particle density $n_{0}=2|\phi_{1s}|^2$ and
 kinetic energy $T_{0}=2t_{1s}$. States 1-4 all have the same density
 $n_{\FE}=|\phi_{1s}|^2+|\phi_{2s}|^2$ and kinetic energy
 $T_{\FE}=t_{1s}+t_{2s}$. Here, $t_i=\ibraketop{i}{\hat{t}}{i}$
 is the single-particle kinetic energy of orbital $i$.
 We see that $\ibraketop{\FE}{\nh}{\FE'}$ and 
 $\ibraketop{\FE}{\Th}{\FE'}$ are thus diagonal, with the exception
 of some elements which mix state 0 with states 1--4.
 We can therefore write
 \begin{align}
  \Gammah_{\mU}=&w_0\iout{0}
  + \sum_{\FE>0}w_{\FE}\iout{\FE_{\mU}}\in \FG^{n,0}
 \end{align}
 where $\iket{\FE_{\mU}}=\sum_{1\leq\FE'\leq 4} U_{\FE'\FE}\iket{\FE'}$ for unitary
 matrix $\mU$ on indices $1\leq \FE,\FE'\leq 4$.

 For the groundstate, we find
 $\Lambda_{\Hx,0}\equiv \ibraketop{0}{\Wh}{0}=\ieri{1s1s}{1s1s}$.
 We now determine the block matrix
 $\ibraketop{\FE}{\Wh}{\FE'}$ for $1\leq \FE,\FE'\leq 4$.
 The Slater-Condon rules give
 $W_{11}=W_{22}=\ieri{1s1s}{2s2s}-\ieri{1s2s}{1s2s}\equiv X-Y$
 and $W_{33}=W_{44}=\ieri{1s1s}{2s2s}\equiv X$. The only non-zero
 cross-terms are
 $\ibraketop{3}{\Wh}{4}=\ibraketop{4}{\Wh}{3}
 =-\ieri{1s2s}{1s2s}=-Y$. Here $\ieri{ij}{kl}
 =\half\int d\vr d\vrp
 \phi^*_i(\vr)\phi_j(\vr)\phi_k^*(\vrp)\phi_l(\vrp) /|\vr-\vrp|$
 is the usual two-electron-repulsion integral.
 Note that $0\leq Y< X$.
 For this block, we find a three-fold degenerate triplet
 eigenvalue $\Lambda_{\Hx,1T}=X-Y$
 (with eigenstates $\iket{\bar{1}/\bar{2}}
 =\iket{1s\up 2s\up}/\iket{1s\down 2s\down}$
 and $\iket{\bar{3}}=[\iket{1s\up 2s\down}
 +\iket{1s\down 2s\up}]/\sqrt{2}$)
 and a higher energy singlet eigenvalue $\Lambda_{\Hx,2S}=X+Y$
 (with eigenstate $\iket{\bar{4}}=[\iket{1s\up 2s\down}
 -\iket{1s\down 2s\up}]/\sqrt{2}$).
 Eq.~\eqref{eqn:EHxLimEig} then gives
 \begin{align}
  \EHx[n]=&w_0\Lambda_{\Hx,0} + w_T\Lambda_{\Hx,1T} + w_4\Lambda_{\Hx,2S}
 \end{align}
 where $w_T=w_1+w_2+w_3$ and $w_4=1-w_0-w_T$.
 Finally, taking derivatives with respect to excited state
 weights $w_T$ and $w_4$ as per \rcite{Pribram-Jones2014} shows that
 a qualitatively correct excitation spectra is obtained.
 By contrast, working directly at the level of $\tr[\Gammah\Wh]$
 within the restriction of single Slater determinants, can lead to
 alternative qualitatively incorrect
 result\cite{Pribram-Jones2014}. This case is
 \emph{automatically excluded} by working within our setting.

 We shall now proceed to work through another case:
 \emph{fully dissociated H$_2$}, exemplifying important size-consistency 
 considerations\cite{GoriGiorgi2008,GoriGiorgi2009}.
 Unlike the He case above, we will this time let the weights
 vary under two constraints: i) the ensemble has only groundstates,
 ii) neither atom has a magnetic moment.
 In an H$_2$ molecule at finite spacing, the four lowest KS states are%
 \footnote{We ignore states with identical spins, inclusion of which 
 complicates details without altering conclusions.}
 \begin{align*}
  \iket{0}=\iket{g\up g\down},
  ~
  \iket{1}=\iket{u\up g\down},
  ~
  \iket{2}=\iket{g\up u\down},
  ~
  \iket{3}=\iket{u\up u\down}.
 \end{align*}
 Here the \emph{gerade}/\emph{ungerade} molecular orbitals $\iket{g/u} =
 \frac{1}{\sqrt{2}} \left[ \iket{a} \pm \iket{b} \right]$ are built
 from localized atomic orbitals, $\iket{a}$ and $\iket{b}$, on atoms
 ``a'' and ``b'', respectively.
 The states $\iket{\FE}$, with $\FE = 0,1,2,3$ have been ordered
 according to their \emph{non}-dissociated energies but are degenerate
 in the dissociated limit considered here. 
 The constraints allow a family of non-interacting
 ``molecular'' ensemble density matrices,
 \begin{align}
  \Gammahmol=&w_0\iout{0} + w_1\iout{1} + w_2\iout{2} + w_3\iout{3}\;,
 \end{align}
 where $\sum w_{\FE}=1$.
 Alternatively, we may take superpositions
 $\iket{\bar{0}/\bar{1}}=[\iket{0}\pm\iket{1}\mp\iket{2}-\iket{3}]/2$
 of Kohn-Sham determinants $\iket{0}$--$\iket{3}$ to obtain localized states,
 \begin{align*}
  \iket{\bar{0}}=\iket{a\up b\down},
  ~
  \iket{\bar{1}}=\iket{b\up a\down},
 \end{align*}
 directly, and thus define a ``localized'' ensemble
 \begin{align}
  \Gammahloc =& \half \iout{\bar{0}} + \half \iout{\bar{1}}\;.
 \end{align}

 Both $\Gammahmol$ and $\Gammahloc$ 
 are in $\FG^{n,0}$ because they obey the constraints,
 provide the same ensemble particle
 density $n= n_a + n_b$ and have the same ensemble kinetic energy
 $\Ts = \sum_{\FE}w_{\FE}T_{\FE}=2\ibraketop{a}{\hat{t}}{a}=2t_a$.
 But their action on the interaction operator $\Wh$ is very different,
 leading to the non-uniqueness disaster if $\tr[\Gammah\Wh]$ is applied
 directly.
 The result $\ibraketop{\FE}{\Wh}{\FE}= \ieri{aa}{aa}$ for all $\FE$
 means that $\tr[\Gammahmol\Wh]=\ieri{aa}{aa}$. By contrast,
 $\ibraketop{\bar{0}}{\Wh}{\bar{0}}=\ibraketop{\bar{1}}{\Wh}{\bar{1}}=0$
 gives $\tr[\Gammahloc \Wh]=0$. It therefore follows that,
 \begin{align}
  \EHx[n] 
  =\tr[\Gammahloc \Wh]= 0\;.
 \end{align}
 The main point here is the following: unless $\FG^{n,0}$ is
 restricted to particular ensembles (e.g. $\Gammahmol$), $\EHx[n]$ is
 not affected by spurious interactions and, thus, size consistency at
 full dissociation is satisfied.

 Finally, it should by now be apparent that we can summarize all
 previous results by writing
 \begin{align}
  \EHx[n] = \min_{ \Gammah \in \FG^{n,\lambda=0} } \tr[\Gammah \Wh]\;,
  \label{eqn:EHxMin3}
 \end{align}
 where we admit into $\FG^{n,\lambda=0}$ all non-interacting
 ensembles of \emph{general form} that simultaneously give the
 prescribed ensemble particle density $n$ and \emph{exact} Kohn-Sham
 kinetic energy $\Ts[n]$, and obey any additional constraints $\FC$.


 To conclude, in this work we have presented a `Hartree-exchange'
 functional [Eq.~\eqref{eqn:EHxLim}] that is uniquely defined in
 ensemble density functional theory for a given set of constraints.
 In a more convenient form [Eq.~\eqref{eqn:EHxMin}],
 it obviously avoids the non-uniqueness pathology, while preserving
 a maximal freedom from spurious interactions.
 The resulting `Hartree-exchange' functional $\EHx[n]$ reproduces
 special cases previously reported in the literature, including
 those requiring non-trivial superpositions of Kohn-Sham states
 [Eq.~\eqref{eqn:EHxLimU}], but can nonetheless straightforwardly
 be obtained via block diagonalisation [Eq.~\eqref{eqn:EHxLimEig}].

 This work will thus aid
 in the development of future EDFT approximations along the lines of
 those previously considered%
 \cite{Nagy1995-exact,Tasnadi2003,Pittalis2006,Perdew2007-OpenX,%
 Gould2013-LEXX,Gould2013-Aff,Gould2014-KS,%
 Fabiano2014,Pribram-Jones2014,Yang2014,%
 Kraisler2013,Kraisler2014,Kraisler2015}%
 , and will extend DFT to new physics.
 The approach can be applied to range-separated interactions
 $\Wh\to \Wh^{\text{sr/lr}}$ to allow for cancellation of errors
 when combining with semi-local approximations.
 Eq.~\eqref{eqn:EHxLimEig} should help in tackling Fermionic
 systems where many states have the same density and kinetic energy
 (e.g. Hubbard models or cold atoms).
 With suitable generalization, this work may be extended to
 open systems, by noting that Eq. \eqref{eqn:EHxLim} depends only
 on the existence and continuity of $\F^{\lambda}[n]$ at $\lambda=0$.
 Generalizations and approximations for the Hx
 and correlation functionals will be presented in future works 
 -- in particular, it will be interesting to gauge the importance
 of ghost-interaction leftovers.

 \acknowledgments

 T.G. would like to thank Leeor Kronik, Andreas Savin and Julien Toulouse
 for helpful discussion and important insights.
 S.P. acknowledges support by the European Community through the
 FP7's Marie-Curie International-Incoming Fellowship, Grant agreement
 No. 623413.

%

\newpage
\hphantom{x}
\includepdf[pages=1]{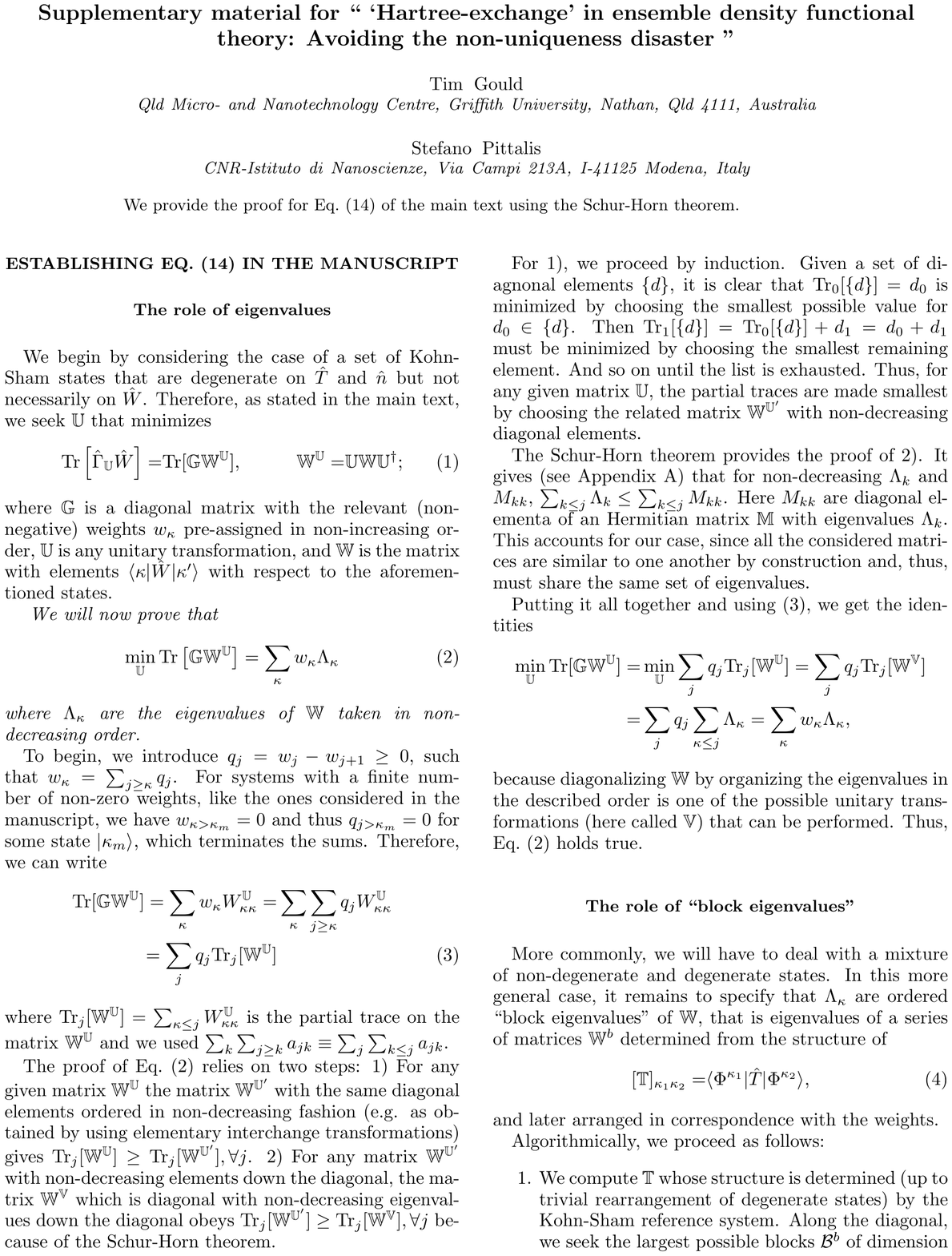}
\newpage
\hphantom{x}


\section*{Supplementary material}

# Supplementary material for " 'Hartree-exchange' in ensemble density functional theory: Avoiding the non-uniqueness disaster "

Tim Gould

*Qld Micro- and Nanotechnology Centre, Griffith University, Nathan, Qld 4111, Australia*

Stefano Pittalis

*CNR-Istituto di Nanoscienze, Via Campi 213A, I-41125 Modena, Italy*

We provide the proof for Eq. (14) of the main text using the Schur-Horn theorem.

## ESTABLISHING EQ. (14) IN THE MANUSCRIPT

### The role of eigenvalues

We begin by considering the case of a set of Kohn-Sham states that are degenerate on $\hat{T}$ and $\hat{n}$ but not necessarily on $\hat{W}$. Therefore, as stated in the main text, we seek $\mathbb{U}$ that minimizes

$$\text{Tr}\left[\hat{\Gamma}_{\mathbb{U}}\hat{W}\right] = \text{Tr}[\mathbb{G}\mathbb{W}^{\mathbb{U}}], \qquad \mathbb{W}^{\mathbb{U}} = \mathbb{U}\mathbb{W}\mathbb{U}^{\dagger}; \qquad (1)$$

where $\mathbb{G}$ is a diagonal matrix with the relevant (non-negative) weights $w_{\kappa}$ pre-assigned in non-increasing order, $\mathbb{U}$ is any unitary transformation, and $\mathbb{W}$ is the matrix with elements $\langle \kappa | \hat{W} | \kappa' \rangle$ with respect to the aforementioned states.

*We will now prove that*

$$\min_{\mathbb{U}} \text{Tr}\left[\mathbb{G}\mathbb{W}^{\mathbb{U}}\right] = \sum_{\kappa} w_{\kappa} \Lambda_{\kappa} \qquad (2)$$

*where $\Lambda_{\kappa}$ are the eigenvalues of $\mathbb{W}$ taken in non-decreasing order.*

To begin, we introduce $q_j = w_j - w_{j+1} \geq 0$, such that $w_{\kappa} = \sum_{j \geq \kappa} q_j$. For systems with a finite number of non-zero weights, like the ones considered in the manuscript, we have $w_{\kappa > \kappa_m} = 0$ and thus $q_{j > \kappa_m} = 0$ for some state $|\kappa_m\rangle$, which terminates the sums. Therefore, we can write

$$\text{Tr}[\mathbb{G}\mathbb{W}^{\mathbb{U}}] = \sum_{\kappa} w_{\kappa} W^{\mathbb{U}}_{\kappa\kappa} = \sum_{\kappa} \sum_{j \geq \kappa} q_j W^{\mathbb{U}}_{\kappa\kappa}$$

$$= \sum_j q_j \text{Tr}_j[\mathbb{W}^{\mathbb{U}}] \qquad (3)$$

where $\text{Tr}_j[\mathbb{W}^{\mathbb{U}}] = \sum_{\kappa \leq j} W^{\mathbb{U}}_{\kappa\kappa}$ is the partial trace on the matrix $\mathbb{W}^{\mathbb{U}}$ and we used $\sum_k \sum_{j \geq k} a_{jk} \equiv \sum_j \sum_{k \leq j} a_{jk}$.

The proof of Eq. (2) relies on two steps: 1) For any given matrix $\mathbb{W}^{\mathbb{U}}$ the matrix $\mathbb{W}^{\mathbb{U}'}$ with the same diagonal elements ordered in non-decreasing fashion (e.g. as obtained by using elementary interchange transformations) gives $\text{Tr}_j[\mathbb{W}^{\mathbb{U}}] \geq \text{Tr}_j[\mathbb{W}^{\mathbb{U}'}], \forall j$. 2) For any matrix $\mathbb{W}^{\mathbb{U}'}$ with non-decreasing elements down the diagonal, the matrix $\mathbb{W}^{\mathbb{V}}$ which is diagonal with non-decreasing eigenvalues down the diagonal obeys $\text{Tr}_j[\mathbb{W}^{\mathbb{U}'}] \geq \text{Tr}_j[\mathbb{W}^{\mathbb{V}}], \forall j$ because of the Schur-Horn theorem.

For 1), we proceed by induction. Given a set of diagonal elements $\{d\}$, it is clear that $\text{Tr}_0[\{d\}] = d_0$ is minimized by choosing the smallest possible value for $d_0 \in \{d\}$. Then $\text{Tr}_1[\{d\}] = \text{Tr}_0[\{d\}] + d_1 = d_0 + d_1$ must be minimized by choosing the smallest remaining element. And so on until the list is exhausted. Thus, for any given matrix $\mathbb{U}$, the partial traces are made smallest by choosing the related matrix $\mathbb{W}^{\mathbb{U}'}$ with non-decreasing diagonal elements.

The Schur-Horn theorem provides the proof of 2). It gives (see Appendix A) that for non-decreasing $\Lambda_k$ and $M_{kk}$, $\sum_{k \leq j} \Lambda_k \leq \sum_{k \leq j} M_{kk}$. Here $M_{kk}$ are diagonal elementa of an Hermitian matrix $\mathbb{M}$ with eigenvalues $\Lambda_k$. This accounts for our case, since all the considered matrices are similar to one another by construction and, thus, must share the same set of eigenvalues.

Putting it all together and using (3), we get the identities

$$\min_{\mathbb{U}} \text{Tr}[\mathbb{G}\mathbb{W}^{\mathbb{U}}] = \min_{\mathbb{U}} \sum_j q_j \text{Tr}_j[\mathbb{W}^{\mathbb{U}}] = \sum_j q_j \text{Tr}_j[\mathbb{W}^{\mathbb{V}}]$$

$$= \sum_j q_j \sum_{\kappa \leq j} \Lambda_{\kappa} = \sum_{\kappa} w_{\kappa} \Lambda_{\kappa},$$

because diagonalizing $\mathbb{W}$ by organizing the eigenvalues in the described order is one of the possible unitary transformations (here called $\mathbb{V}$) that can be performed. Thus, Eq. (2) holds true.

### The role of "block eigenvalues"

More commonly, we will have to deal with a mixture of non-degenerate and degenerate states. In this more general case, it remains to specify that $\Lambda_{\kappa}$ are ordered "block eigenvalues" of $\mathbb{W}$, that is eigenvalues of a series of matrices $\mathbb{W}^b$ determined from the structure of

$$[\mathbb{T}]_{\kappa_1 \kappa_2} = \langle \Phi^{\kappa_1} | \hat{T} | \Phi^{\kappa_2} \rangle, \qquad (4)$$

and later arranged in correspondence with the weights.

Algorithmically, we proceed as follows:

1. We compute $\mathbb{T}$ whose structure is determined (up to trivial rearrangement of degenerate states) by the Kohn-Sham reference system. Along the diagonal, we seek the largest possible blocks $\mathcal{B}^b$ of dimension

$N^b \times N^b$ for which $T_{\kappa_1 \kappa_2} = T^b \delta_{\kappa_1 \kappa_2}$ is a constant $T^b$ times the identity matrix within the block. *Note, the ordering of these blocks must be preserved in subsequent steps.*

We illustrate such a matrix below for the case of three blocks,

$$\mathbb{T} = \begin{pmatrix} T_{00} & \times & \times & \times & \times & \times & \times \\ \times & T_{11} & 0 & 0 & 0 & \times & \times \\ \times & 0 & T_{22} & 0 & 0 & \times & \times \\ \times & 0 & 0 & T_{33} & 0 & \times & \times \\ \times & 0 & 0 & 0 & T_{44} & \times & \times \\ \times & \times & \times & \times & \times & T_{55} & 0 \\ \times & \times & \times & \times & \times & 0 & T_{66} \end{pmatrix}.$$

Here the three blocks $\mathcal{B}^1$, $\mathcal{B}^2$ and $\mathcal{B}^3$ have $N^1 = 1$, $N^2 = 4$ and $N^3 = 2$ with block kinetic energies $T^1 = T_{00}$, $T^2 = T_{11} = T_{22} = T_{33} = T_{44}$ and $T^3 = T_{55} = T_{66}$.

2. Next, we identify the corresponding blocks $\mathcal{B}^b$ in $\mathbb{W}$ and calculate the "block eigenvalues" $\Lambda_\xi^b$ of the matrix $\mathbb{W}^b$ defined on elements in $\mathcal{B}^b$ only. We label these eigenvalue $\Lambda_\xi$ by an index $0 \leq \xi < N^b$, and order them non-decreasingly, within each block.

We show below the blocks $\mathbb{W}^1$, $\mathbb{W}^2$ and $\mathbb{W}^3$ corresponding to $\mathbb{T}$ above,

$$\mathbb{W} = \begin{pmatrix} \mathbb{W}^1 & \times & \times & \times & \times & \times & \times \\ \times & & & & & \times & \times \\ \times & & \mathbb{W}^2 & & & \times & \times \\ \times & & & & & \times & \times \\ \times & & & & & \times & \times \\ \times & \times & \times & \times & \times & \mathbb{W}^3 & \\ \times & \times & \times & \times & \times & & \end{pmatrix}.$$

These are then diagonalized separately to obtain $\{\Lambda_0^1\}$, $\{\Lambda_0^1, \Lambda_1^1, \Lambda_2^1, \Lambda_3^1\}$ and $\{\Lambda_0^3, \Lambda_1^3\}$, respectively for $\mathbb{W}^1$, $\mathbb{W}^2$ and $\mathbb{W}^3$.

3. Finally, taking the blocks in their original order, we perform the weighted sum

$$\mathcal{E}_{\text{Hx}}[n] = \sum_b \sum_{0 \leq \xi < N^b} w_{\kappa^b + \xi} \Lambda_\xi^b \quad (5)$$

where $\kappa^b = \sum_{b' < b} N^{b'}$ labels the first element of block $\mathcal{B}^b$, and $\Lambda_\xi^b$ is the $\xi$th eigenvalue of $\mathbb{W}^b$. Here we use the results from the previous section within each block (where degenerate perturbation theory allows unitary transformations), but not between blocks (which non-degenerate perturbation theory forbids mixing). *It is important to note that neighbouring elements from different blocks can decrease in value even if $\kappa$ increases.*

Finally, from Eq. (5) we arrive at Eq. (14) of the main text by simply stressing the energy meaning of the aforementioned block eigenvalues as well as their functional dependency on the overall ensemble particle density

$$\mathcal{E}_{\text{Hx}}[n] = \sum_\kappa w_\kappa \Lambda_{\text{Hx},\kappa}[n] \quad (6)$$

i.e., we have renamed/relabeled $\kappa^b + \xi \to \kappa$ and $\Lambda_\xi^b \to \Lambda_{\text{Hx},\kappa}[n]$. As elsewhere, all functionals of $n$ also depend on our choice of constraints $\mathcal{C}$.

In our example this is,

$$\begin{aligned}\mathcal{E}_{\text{Hx}}[n] =& [w_0 \Lambda_0^1] \\ &+ [w_1 \Lambda_0^2 + w_2 \Lambda_1^2 + w_3 \Lambda_2^2 + w_4 \Lambda_3^2] \\ &+ [w_5 \Lambda_0^3 + w_6 \Lambda_1^3] \\ \equiv& \sum_\kappa w_\kappa \Lambda_{\text{Hx},\kappa}[n].\end{aligned}$$

Here, $\Lambda_0^2$ can potentially be smaller than $\Lambda_0^1$ (similarly for $\Lambda_0^3$ and $\Lambda_3^2$) due to the block structure.

### Appendix A: Schur-Horn corrollary

Commonly, the Schur-Horn theorem is written with partial sums of non-increasing elements which obey

$$\sum_{k \leq j} M_{kk} \leq \sum_{k \leq j} \Lambda_k , \quad \sum_{k \leq N} M_{kk} = \sum_{k \leq N} \Lambda_k = \text{Tr}[\mathbb{M}] ,$$

where $M_{kk}$ are the diagonal elements of matrix $\mathbb{M}$ and $\Lambda_k$ are its eigenvalues. See e.g. the Wikipedia entry at https://en.wikipedia.org/wiki/Schur\%E2\%80\%93Horn_theorem.

The version employed here is a direct corrollary, as shall be shown below. Firstly, we set $M'_{kk} = M_{(N-k)(N-k)}$ and $\Lambda'_k = \Lambda_{N-k}$ to reverse the order so that elements are non-decreasing. This gives,

$$\sum_{j' < k \leq N} M'_{kk} \leq \sum_{j' < k \leq N} \Lambda'_k , \quad \sum_{k \leq N} M'_{kk} = \sum_{k \leq N} \Lambda'_k = \text{Tr}[\mathbb{M}] ,$$

for non-decreasing elements and $j' + 1 = N - j$.

We then write $\sum_{k \leq j} M_{kk} + \sum_{j < k \leq M} M_{kk} = \sum_{k \leq N} M_{kk} = \text{Tr}[\mathbb{M}]$ and $\sum_{k \leq j} \Lambda_k + \sum_{j < k \leq N} \Lambda_k = \sum_{k \leq N} \Lambda_k = \text{Tr}[\mathbb{M}]$ (where we drop primes and used the invariance of the trace). Therefore,

$$0 \geq \sum_{j < k \leq N} M_{kk} - \sum_{j < k \leq N} \Lambda_k = \sum_{k \leq j} \Lambda_k - \sum_{k \leq j} M_{kk}$$

and, finally, we get the desired

$$\sum_{k \leq j} \Lambda_k \leq \sum_{k \leq j} M_{kk} .$$



\end{document}